\documentclass[aps,twocolumn,preprintnumbers,nofootinbib,superscriptaddress]{revtex4-1}
\usepackage{amsmath, amssymb} \usepackage{graphicx} \usepackage{amsfonts}
\usepackage{array} \usepackage{amsthm} \usepackage{bm}
\usepackage{palatino} \usepackage{mathpazo} 
\usepackage{supertabular}

\usepackage[breaklinks]{hyperref}
\usepackage{color}

\newcommand{\nn}{\nonumber}
\newcommand{\be}{\begin{equation}}
\newcommand{\ee}{\end{equation}}
\newcommand{\ba}{\begin{eqnarray}}
\newcommand{\ea}{\end{eqnarray}}
\newcommand{\bal}{\begin{align}}
\newcommand{\eal}{\end{align}}

\newcommand{\e}{{\rm e}}
\newcommand{\dd}{{\rm d}}

\newcommand{\bb}{\bibitem}

\newcommand{\al}{\alpha}

\newcommand{\bt}{\beta}

\newcommand{\bw}{\begin{widetext}}
\newcommand{\ew}{\end{widetext}}

\begin{document}
\title{Smoothed one-core and core--multi-shell regular black holes}

\author{Mustapha Azreg-A\"{\i}nou}
\affiliation{Ba\c{s}kent University, Engineering Faculty, Ba\u{g}l\i ca Campus, Ankara 06810, Turkey}


\begin{abstract}
We discuss the generic properties of a general, smoothly varying, spherically symmetric mass distribution $\mathcal{D}(r,\theta)$, with no cosmological term  ($\theta$ is a length scale parameter). Observing these constraints, we show that (a) the de Sitter behavior of spacetime at the origin is generic and depends only on $\mathcal{D}(0,\theta)$, (b) the geometry may posses up to $2(k+1)$ horizons depending solely on the total mass $M$ if the cumulative distribution of $\mathcal{D}(r,\theta)$ has $2k+1$ inflection points, and (c) no scalar invariant nor a thermodynamic entity diverges. We define new two-parameter mathematical distributions mimicking Gaussian and step-like functions and reduce to the Dirac distribution in the limit of vanishing parameter $\theta$. We use these distributions to derive in closed forms asymptotically flat, spherically symmetric, solutions that describe and model a variety of physical and geometric entities ranging from noncommutative black holes, quantum-corrected black holes to stars and dark matter halos for various scaling values of $\theta$. We show that the mass-to-radius ratio $\pi c^2/G$ is an upper limit for regular-black-hole formation. Core--multi-shell and multi-shell regular black holes are also derived.
\end{abstract}


\maketitle

\section{\hspace{-2mm}Distributions smoothing out the Dirac's $\,\pmb{\delta}$\label{seci}}
One- and multi-parameter-dependent mathematical distributions smoothing out the Dirac's $\delta$ distribution are needed in areas of science where the notion of locality is being abandoned. For instance, in quantum gravity the noncommutativity of coordinates is phenomenologically explained by the nonlocality of matter distributions~\cite{SS3}. The singularities arising in classical physics are due to the hypothetical point-like matter distributions. Such a point-like or Dirac distribution is mathematically useful in getting closed-form simple expressions for the physical and geometric entities one is concerned with. In a sense, the Schwarzschild, Reissner-Nordstr\"om, Kerr and other classical solutions of general relativity are extremely simplified models of nature and should exist in a real world only asymptotically. In some other instances of science, as is the case with regular black holes sourced by nonlinear electrodynamics, such distributions were not needed. That remains true, however, as far as one is concerned with macroscopic scales; for scales of the order of the Compton wavelength or the Planck length, the contribution of the vacuum, namely its radial pressure sustaining matter from collapsing, renders mass distributions extended.

In a first tentative one may think to replace the Dirac distribution for mass by a central---decreasing as one moves away from the source---extended distribution. For spherically symmetric solutions, a Gaussian mass distribution with width $\theta$,
\begin{equation}\label{i1}
\mathcal{G}(r,\theta)= \frac{\e^{-r^2/(2\theta^2)}}{(2\pi)^{3/2}\theta^3},
\end{equation}
where $r$ is a radial coordinate, satisfies the above-mentioned requirement. However, the resulting metric and fields are not in closed-forms and are not easily handled numerically---not to mention analytically---via computer algebra systems~\cite{SS3}. Do extended distributions for charge and spin (if the solution is rotating) follow the same mass-distribution model? In Ref.~\cite{comment} noncommutative charged black holes, with a Gaussian charge distribution, were determined, while in Ref.~\cite{Weibull} it was argued that, if masses follow Gaussian distributions, charges could, rather, follow extended Weibull distributions to ensure a de Sitter behavior of the solution in the vicinity of the origin and noncommutative charged black holes, with a Weibull charge distribution, were determined.

Whether the gravitational quantum effects are well understood or not, introducing them phenomenologically via mass, charge, and spin distribution functions seems to be a fruitful way as this cures singularities, skips the matching problems, and preserves the asymptotical behavior. There remains to understand how the vacuum responds to the mass, charge, and spin extended distributions to generate negative pressures sustaining matter from collapsing. The only known process to advance an explanation for that is vacuum fluctuations but so far no concrete formulation seems to exist.

For the Gaussian distribution the substitution rule Dirac-to-Gaussian reads
\begin{equation}\label{sr1}
\frac{\delta(r)}{2\pi r^2}\to \frac{\e^{-r^2/(2\theta^2)}}{(2\pi)^{3/2}\theta^3},
\end{equation}
where the numerical coefficients in~\eqref{i1} and~\eqref{sr1} have been determined on observing the normalization conditions
\[\int_0^{\infty}\frac{\delta(r)}{2\pi r^2}~4\pi r^2\dd r=\int_0^{\infty}\frac{\e^{-r^2/(2\theta^2)}}{(2\pi)^{3/2}\theta^3}~4\pi r^2\dd r=1.\]

Let $\mathcal{D}(r,\theta)\geq 0$ be some spherically symmetric, not necessarily central, distribution with the normalization condition
\begin{equation}\label{i2}
\int_0^{\infty}\mathcal{D}(r,\theta)~4\pi r^2\dd r=1.
\end{equation}
If $\mathcal{D}$ is a mass distribution one may think of it to be central, however, the vacuum negative radial pressure, too assumed to be spherically symmetric, may push more matter from the center rendering the distribution non-central with voids. This is the case with four-dimensional charge distributions~\cite{Weibull}, which may be non-central (of Weibull character in some instances) and vanish at the origin. Three-dimensional non-central mass distributions with a central void have been shown to exist too~\cite{void}. The presence of the central void is to ensure existence of a two-horizon structure.

Let $m(r,\theta)$ denote the mass inside a sphere of radius $r$. This is given by
\begin{equation}\label{i3}
m(r,\theta)=M\int_0^{r}\mathcal{D}(r',\theta)4\pi r'^2\dd r'=M\mathbb{D}(r,\theta),
\end{equation}
where $M$ is the total mass of the solution. To simplify the notation we have set
\begin{equation}\label{i3b}
\mathbb{D}(r,\theta)\equiv \int_0^{r}\mathcal{D}(r',\theta)4\pi r'^2\dd r',
\end{equation}
which is the cumulative distribution. The above substitution rule~\eqref{sr1} is replaced by
\begin{equation}\label{sr2}
\frac{\delta(r)}{2\pi r^2}\to \mathcal{D}(r,\theta).
\end{equation}

The constraint~\eqref{i2} implies that $\mathbb{D}\to 1$ as $r\to\infty$. This excludes from our analysis the de Sitter-like solutions\footnote{A de Sitter-like metric includes a term proportional to $-r^2$, which can be arranged as $-m(r)/r$ with $m(r)\propto r^3$ [compare with~\eqref{i4}]. This means that $m(r)$ and $\mathbb{D}(r)$ go to $\infty$ as $r\to \infty$.}, as those treated in~\cite{entropy}, where $\mathbb{D}\to \infty$ as $r\to\infty$, and the anti-de Sitter-like solutions, where $\mathbb{D}$ turns negative for some $r>0$.

It is understood that the distribution $\mathcal{D}(r,\theta)$, smoothing out the Dirac's one, is assumed to be finite and differentiable for all $r$. This implies that $\partial_r\mathcal{D}(r,\theta)\equiv \mathcal{D}'(r,\theta)$ has finite values for all $r$. The convergence of the integral in~\eqref{i2} implies that $\mathcal{D}$ must go to 0 faster than $1/r^3$ in the limit $r\to\infty$. These requirements are expressed mathematically as
\begin{align}
\label{sr3a}&0<\mathcal{D}(r,\theta)<\infty,\\
\label{sr3b}&-\infty<\mathcal{D}'(r,\theta)<\infty,\\
\label{sr3c}&\lim_{r\to \infty}r^3\mathcal{D}=0.
\end{align}

In Sec.~\ref{secp} we discuss the generic properties of any mass distribution obeying~\eqref{sr3a}-\eqref{sr3c} and of its resulting metric solution. In Sec.~\ref{secd} we define new two-parameter, ($n,\,\theta$), mathematical distributions mimicking the Gaussian distribution and reduce to the Dirac distribution in the limit of vanishing parameter $\theta$ (for all $n$) and discuss their specific properties and the properties of their resulting metric solutions. In Sec.~\ref{seclc} we discuss some limiting cases. In Sec.~\ref{secs} we provide instances of applications ranging from noncommutative black holes, quantum-corrected black holes to stars and dark matter halos for various scaling values of $\theta$. An Appendix section has been added to complete the discussion of, and to derive some equations pertaining to, Sec.~\ref{secp}. We conclude in Sec.~\ref{secc}.

\section{Generic properties of the metric\label{secp}}

We seek a static spherically symmetric solution of the form
\begin{align}\label{i4}
&\dd s^2=f(r)\dd t^2-\frac{\dd r^2}{f(r)}-r^2(\dd\vartheta^2+\sin^2\vartheta \dd\varphi^2),\\
&f(r)=1-\frac{2Gm(r,\theta)}{c^2r}=1-\frac{2MG\,\mathbb{D}(r,\theta)}{c^2r},\nn
\end{align}
where $m(r,\theta)$ is given by~\eqref{i3}. In the following we discuss the generic properties of~\eqref{i4} for a matter distribution obeying the minimum set of constraints~\eqref{sr3a}, \eqref{sr3b}, and ~\eqref{sr3c}.\\

\paragraph{\textbf{Behavior near the origin.---}}

Since $\mathcal{D}(r,\theta)$ is assumed to be finite everywhere, for $r\ll 1$, we may replace $\mathcal{D}(r',\theta)$ in~\eqref{i3} by $\mathcal{D}(0,\theta)$ to obtain
\begin{equation}\label{i5}
m(r,\theta)\underset{(r\to 0)}{\simeq} M\int_0^{r}\mathcal{D}(0,\theta)4\pi r'^2\dd r'=
\frac{4\pi M\mathcal{D}(0,\theta)}{3}~r^3,
\end{equation}
yielding
\begin{equation}
f\underset{(r\to 0)}{\simeq}1-\frac{8\pi MG\mathcal{D}(0,\theta)}{3c^2}~r^2.
\end{equation}
Thus, any distribution with nonvanishing value at the origin [$\mathcal{D}(0,\theta)\neq 0$] yields a metric having a de Sitter behavior there with an effective ``cosmological constant"
\begin{equation}\label{ic}
\Lambda=8\pi MG\mathcal{D}(0,\theta)/c^2,
\end{equation}
linearly proportional to the total mass $M$ as far as the width $\theta$ does not dependent on the mass. The distribution need not be central to yield such a behavior for $f$: All that we need is to have $\mathcal{D}(0,\theta)\neq 0$.\\

\paragraph{\textbf{The scalar invariants.---}}

With the further assumption that $\mathcal{D}'(r,\theta)$ has a finite value at $r=0$~\eqref{sr3b} and that $\mathcal{D}(r,\theta)$ executes smooth variations near the origin, it is straightforward to show that the curvature and Kretschmann scalars are finite at the origin:
\begin{align}
\label{cs1}\mathcal{R}=& -\frac{8\pi MG}{c^2} ~(4\mathcal{D}+r\mathcal{D}'),\\
\mathcal{R}_{\al\bt\mu\nu}\mathcal{R}^{\al\bt\mu\nu}=&\frac{16G^2 m [3 m+4\pi M r^3 (r \mathcal{D}'-2 \mathcal{D})]}{c^4r^6}\nn\\
&+\frac{64 \pi ^2 M^2G^2}{c^4}~(4 \mathcal{D}^2+r^2 \mathcal{D}'^2).
\end{align}
Since $m(r,\theta)$ behaves as $r^3$~\eqref{i5} near the origin, we see that both expressions of $R$ and $R_{\al\bt\mu\nu}R^{\al\bt\mu\nu}$ have finite limits as $r\to 0$. Thus, the singularity at the origin has been removed. Moreover, since $\mathcal{D}'(r,\theta)$ is finite for all $r$~\eqref{sr3b}, the two scalar invariants remain finite too for all $r>0$.\\

\paragraph{\textbf{Horizons.---}}

\begin{figure*}
\centering
\includegraphics[width=0.24\textwidth]{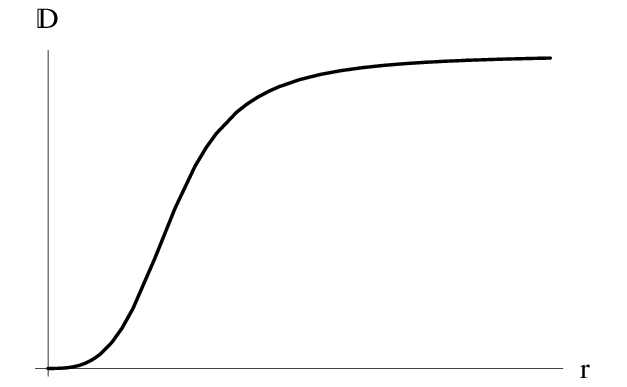} \includegraphics[width=0.24\textwidth]{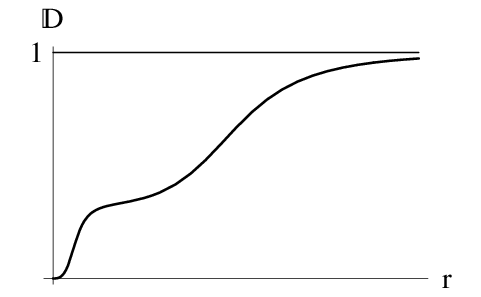} \includegraphics[width=0.24\textwidth]{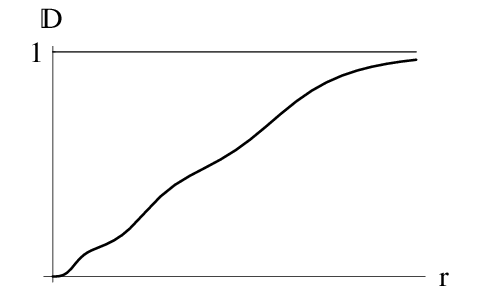} \includegraphics[width=0.24\textwidth]{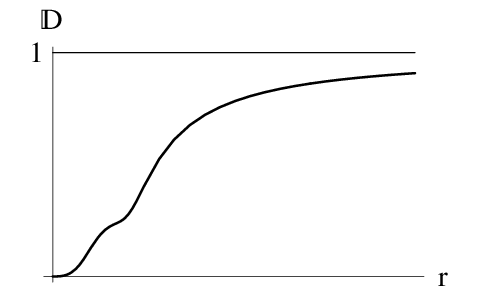} \\
\caption{\footnotesize{Generic plots of the cumulative distribution $\mathbb{D}(r,\theta)$~\eqref{i3b} whether $\theta$ depends on $M$ or not. In the left-most plot $\mathbb{D}(r,\theta)$ has one point of inflection and the corresponding metric may have up to two horizons. In the second plot from the left $\mathbb{D}(r,\theta)$ has three inflection points and the corresponding metric may have up to four horizons [this is the plot of the cumulative distribution of the core--shell regular black hole~\eqref{w4} taking $M_2=2/3=2M_1$ ($M=1$) and $\theta_2=5=8\theta_1$]. In the third plot from the left $\mathbb{D}(r,\theta)$ has five inflection points and the corresponding metric may have up to six horizons [this is the plot of the cumulative distribution of the core--two-shell regular black hole~\eqref{w6} taking $M_1=0.3$, $M_2=0.8$, $M_3=1.2$ ($M=2.3$), $\theta_1=1$, $\theta_2=4$, and $\theta_3=9.5$]. In the right-most plot $\mathbb{D}(r,\theta)$ has three inflection points and the corresponding metric may have up to four horizons [this is the plot of the cumulative distribution of $\mathcal{D}(r,\theta)$ as given by~\eqref{w7}, \eqref{w8} and~\eqref{w9}, taking $\theta_1=\theta_2=a\equiv \theta=1$, $M_1=M/7$, and $M_2=6M/7$. This describes another core--shell regular black hole]. In all these cases the origin has been excluded.}}\label{Fig1}
\end{figure*}

The horizons, all denoted by $r_h$, are solutions to the equation $f(r_h)=0$, which reduces to
\begin{equation}\label{p1}
\frac{c^2}{2MG}~r_h=\mathbb{D}(r_h,\theta),
\end{equation}
where we have used~\eqref{i3} and~\eqref{i3b}. In the $r_hy$ plane, the horizons are the intersection points of the straight line $y=c^2r_h/(2MG)$ and the curve $y=\mathbb{D}(r_h,\theta)$ among which we find the point $r_h=0$, which we exclude. By~\eqref{i5} the graph of $y=\mathbb{D}(r_h,\theta)$ is flat at the origin (in the limit $r_h\to 0$) and by~\eqref{i2} it is also flat asymptotically (in the limit $r_h\to \infty$). Since $\mathbb{D}(r,\theta)$ is the cumulative distribution it is an increasing function of $r$ [$\mathbb{D}'=4\pi r^2\mathcal{D}> 0$~\eqref{sr3a}], so its shape looks like a flat S if it has one inflection point or like a step function with two steps if it has three inflection points (with more steps if it has more than three inflection points), as depicted in Fig.~\ref{Fig1}. This generic graph of $\mathbb{D}(r,\theta)$, as is the case with any cumulative distribution, does not depend on $\theta$ and on whether the latter depends on $M$ or not. It is now clear that for large $M$, the slope of the line $y=c^2r_h/(2MG)$ is small enough to have some intersection points with the curve $y=\mathbb{D}(r_h,\theta)$, that is, up to two horizons $r_h\neq 0$ if $\mathbb{D}(r_h,\theta)$ has one inflection point and up to $2(k+1)$ horizons if $\mathbb{D}(r_h,\theta)$ has $2k+1$ inflection points. As  $M$ decreases and reaches some value $M_{\text{ext}}$, the slope of $y=c^2r_h/(2MG)$ increases until the line becomes tangent to $y=\mathbb{D}(r_h,\theta)$, with no more intersection points, yielding an extremal black hole solution with one horizon $r_{\text{ext}}\neq 0$. For $M<M_{\text{ext}}$ there are no horizons. To summarize, the metric~\eqref{i4} will have the following generic properties:
\begin{widetext}
\begin{align}\label{p2}
&\text{up to $2(k+1)$ horizons if:}& & M>M_{\text{ext}} \text{ and }\mathbb{D} \text{ has $2k+1$ inflection points} & &\text{(black hole solution)},\nn\\
&\text{one horizon if:}& & M=M_{\text{ext}}& &\text{(extremal BH solution)},\nn\\
&\text{no horizon if:}& & M<M_{\text{ext}}& &\text{(particle-like solution)},
\end{align}
\end{widetext}
whether $\theta$ depends on $M$ or not. The $r$ coordinates of the inflection points of $\mathbb{D}(r_h,\theta)$ are solutions to
\begin{equation}\label{ip}
2\mathcal{D}+r\mathcal{D}'=0.
\end{equation}
In Sec.~\ref{paraset} we will show  that the tangential pressure is given by $p_t=-c^2\rho_m-c^2r\partial_r\rho_m/2$ where $\rho_m=M\mathcal{D}$ is the mass density. Equation~\eqref{ip} is just $p_t=0$. This is precisely the equation used in Ref.~\cite{ref2} to determine the number of horizons. This simply implies that at each inflection point of $\mathbb{D}(r_h,\theta)$ the tangential pressure vanishes.

Note that regular multi-horizon solutions are well known objects in the scientific literature (see, for instance, \cite{ref1,ref3} and some other references therein).

Let $x_{\text{ext}}\equiv r_{\text{ext}}/\theta$ and generally
\begin{equation}\label{d0b}
    x\equiv r/\theta.
\end{equation}
We show in the Appendix that $x_{\text{ext}}$ is a solution to
\begin{equation}\label{p3}
\mathcal{D}(x_{\text{ext}})x_{\text{ext}}^3=\int_{0}^{x_{\text{ext}}}\mathcal{D}(u)u^2\dd u,
\end{equation}
and that $M_{\text{ext}}$ is given by
\begin{equation}\label{p3b}
\frac{2M_{\text{ext}}G}{c^2}=\frac{1}{4\pi \theta^2\mathcal{D}(x_{\text{ext}})x_{\text{ext}}^2}.
\end{equation}
We see that if $\mathcal{D}$ does not depend on the mass $M$, this will be the case for $x_{\text{ext}}$ too.

For large, massive black holes one of the nonzero horizons, the inner most horizon $r_{h-}$, shrinks to 0 while the outer most horizon $r_{h+}$ goes to infinity. For the latter horizon we let $r_h\to\infty$, so that the r.h.s of~\eqref{p1} is 1 by~\eqref{i2} implying
\begin{equation}\label{p4}
r_{h+}\underset{(M \text{ large})}{\simeq} r_{\text{S}}\equiv \frac{2MG}{c^2},
\end{equation}
where $r_{\text{S}}$ denotes the Schwarzschild radius.

It is clear from~\eqref{i2} and~\eqref{i3} that the metric~\eqref{i4} is asymptotically flat.\\

\begin{figure}
\centering
\includegraphics[width=0.43\textwidth]{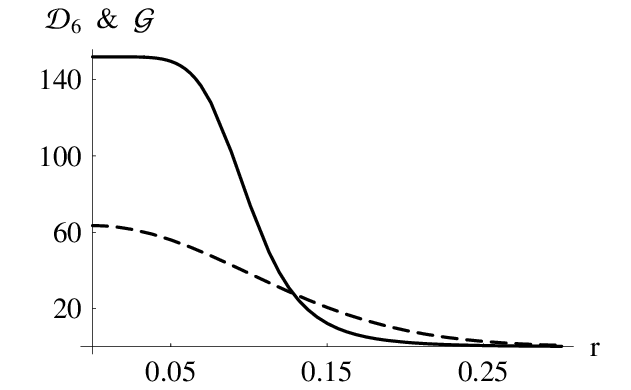} \\
\caption{\footnotesize{The continuous plot represents the step-like $\mathcal{D}_6$ distribution~\eqref{d3} and the dashed plot represents the Gaussian distribution~\eqref{i1} $\mathcal{G}$ for the same value of $\theta=0.1$. Near the black hole $\mathcal{D}_6$, and generally $\mathcal{D}_n$, goes to 0 faster than the Gaussian distribution but far from the horizon this order is reversed (for clarity this is not shown in the plot).}}\label{Fig2}
\end{figure}
\begin{figure}
\centering
\includegraphics[width=0.43\textwidth]{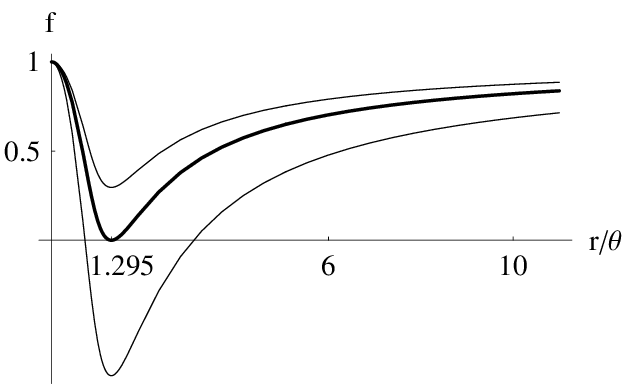} \\
\caption{\footnotesize{Plots of the metric $f$ versus $r/\theta$. Upper plot: A (non-black-hole) quantum particle~\eqref{s1} for $MG/(\pi c^2\theta)=0.2$. Intermediate plot: An extremal noncommutative black hole~\eqref{s1} for $MG/(\pi c^2\theta)=0.284$. Lower plot: A noncommutative black hole~\eqref{s1} with two horizons for $MG/(\pi c^2\theta)=0.5$.}}\label{Fig3}
\end{figure}

\paragraph{\textbf{The temperature.---}}

Due to quantum effects near the event horizon, black holes emit Hawking radiation at the temperature~\cite{Hawking}
\begin{align}
T=&\frac{\hbar c}{4\pi k_{\text{B}}}~\partial_r f\big|_{r=r_{h+}},\nn\\
\label{t1}=&\frac{\hbar c}{4\pi k_{\text{B}}r_{h+}}~\frac{2MG}{c^2}\Big(\frac{\mathbb{D}}{r}-\mathbb{D}'\Big)\Big|_{r=r_{h+}},
\end{align}
where $\hbar$ and $k_{\text{B}}$ are the reduced Planck and Boltzmann constants. In order to investigate the behavior of the temperature as a varying function, it has become customary to express it in terms of the radius of the event horizon $r_{h+}$~\cite{Dym}. If it were possible to express $T$ in terms of the mass $M$ of the black hole, as in the Schwarzschild case, we would have an explicit $T$-$M$ relation. However, since in our case it is not possible to solve~\eqref{p1} for $r_{h+}$ in terms of $M$, all we can do is to express $T$ in terms of $r_{h+}$ as done in~\cite{Dym}.

We obtain $\mathbb{D}'$ upon differentiating~\eqref{i3b} with respect to $r$ and we use~\eqref{p1} to express $(\mathbb{D}/r)|_{r=r_{h+}}$ in terms of $c^2/(2MG)$. Finally, we arrive at
\begin{equation}\label{t2}
    T(r_{h+})=\frac{\hbar c}{4\pi k_{\text{B}}r_{h+}}\big(1-4\pi r_{\text{S}}~r_{h+}^2\mathcal{D}(r_{h+})\big),
\end{equation}
where $r_{\text{S}}=r_{h+}/\mathbb{D}(r_{h+},\theta)$~\eqref{p1} and $T$ is seen as a function of $r_{h+}$.

Using~\eqref{sr3c} and~\eqref{p2}, we see that for large massive black holes $(MG/c^2)r_{h+}^2\mathcal{D}(r_{h+})\to 0$ and
\begin{equation}\label{t3}
T\simeq \frac{\hbar c}{4\pi k_{\text{B}}r_{\text{S}}},
\end{equation}
which is the well-known expression for the Schwarzschild black hole temperature. Note that $r_{\text{ext}}$ is the minimum value of $r_{h+}$. In the Appendix we show that $T(r_{\text{ext}})=0$. Now, since $T$ vanishes at $r_{\text{ext}}$ and it is positive~\eqref{t3} for large values of $r_{h+}$, it must reach some maximum value for some $\tilde{r}_{h+}>r_{\text{ext}}$, then, by~\eqref{t3}, goes to zero as $r_{h+}$ approaches infinity. In the Appendix we show that $\tilde{r}_{h+}$ is solution to
\begin{equation}\label{t4}
\partial_r\Big(\frac{4\pi r^2\mathcal{D}}{\mathbb{D}}\Big)\bigg|_{r=\tilde{r}_{h+}}=-\frac{1}{\tilde{r}_{h+}^2}.
\end{equation}

We see that an evaporation process which starts at some value of $r_{h+}>\tilde{r}_{h+}$ leads, after some loss of matter, to a configuration where the temperature becomes initially larger than the temperature of the starting point, it attains its maximum value at $\tilde{r}_{h+}$, then it drops to zero as $r_{h+}$ reaches the value $r_{\text{ext}}$, which marks the end of the evaporation process for there will be no black hole. The remaining mass is a cold, at $T=0$, regular non-black-hole solution~\eqref{p2}.\\

\paragraph{\textbf{The stress-energy tensor.---}\label{paraset}}

The stress-energy tensor (SET) sourcing the metric~\eqref{i4} is assumed to satisfy $G_{\mu\nu}=(8\pi G/c^4)T_{\mu\nu}$. It has the algebraic structure
\begin{equation}\label{set1}
	T^t_t=T^r_r,\quad T^{\theta}_{\theta}=T^{\varphi}_{\varphi}.
\end{equation}
The resulting equation of state reads
\begin{equation}\label{set2}
p_r=-c^2\rho_m,\quad p_t=-c^2\big(\rho_m+\tfrac{r\partial_r\rho_m}{2}\big),
\end{equation}
where $p_r\equiv -T^r_r$ is the radial pressure and $p_t\equiv -T^{\theta}_{\theta}$ ($\neq p_r$) is the tangential pressure. Here $\rho_m$ is given by
\begin{equation}
\rho_m\equiv \partial_r m(r,\theta)/(4\pi r^2)=M\mathcal{D}(r,\theta).
\end{equation}
The SET~\eqref{set1}, being invariant under boosts in the radial direction (having an infinite set of comoving reference frames), is commonly identified as describing a spherically symmetric anisotropic vacuum~\cite{cosmo3}. This sort of SET belongs to the so-called family of cosmological tensors or variable cosmological term~\cite{cosmo1,cosmo2} where the vacuum state behaves as a de Sitter one, in the vicinity of the origin, and as a Minkowski one, at spatial infinity. A corresponding black hole solution is sometimes called a $\Lambda_{\mu\nu}$BH~\cite{cosmo1}.  

As we shall see in Secs.~\ref{secps} and~\ref{secs} the model, described by~\eqref{set2}, has different scales from cosmological to elementary particles.
In the vicinity of the origin we certainly have $p_t<0$ but its sign may change as $r$ increases. The negative radial pressure is necessary for preventing matter from collapsing and forming a singularity.\\

\paragraph{\textbf{Energy conditions and change of spatial topology.---}\label{paratopo}}

From~\eqref{set2} we see that the Weak Energy Condition (WEC), which requires,
\begin{equation}\label{tp2}
\rho_m\geq 0,\quad c^2\rho_m+p_r\geq 0,\quad c^2\rho_m+p_t\geq 0,
\end{equation}
is violated by non-central mass distributions where $\partial_r\rho_m$ is positive in the vicinity of the center.

Central mass distributions, where $\partial_r\rho_m<0$, do not violate the WEC. Previous studies~\cite{topo1} have shown that under the constraints of the Null Energy Condition (NEC), which are the second and third conditions in~\eqref{tp2}, there may be a change of spatial topology if the black hole is to be regular. Since $\rho_m>0$ for the type of regular black holes we are considering, the WEC and NEC are equivalent. 

More accurately, it was shown that if a black hole spacetime contains trapped surfaces and satisfies the weak energy condition, then there must be a change of spatial topology if the black hole is to be regular~\cite{topo1}. Inside the horizon there is a region where the spatial topology changes from open to compact slices; that is, the spacetime changes its spatial topology from $S^2\times R$ to $S^3$ as the non-compact spacelike three-dimensional hypersurfaces (slices) at spatial infinity evolve to future-trapped compact three-dimensional slices inside the horizon. This can be seen from Figure 1 of Ref.~\cite{topo1} which depicts the conformal global structure of a portion of a regular black hole. In such diagrams each point represents a two-sphere described by two spacelike coordinates, the remaining spacelike coordinate (which runs from $-\infty$ to $+\infty$) is represented horizontally, and the time coordinate is represented vertically. So the line $\mathcal{S}_1$ of Figure 1 of Ref.~\cite{topo1} represents a three-dimensional cylinder which connects two regions at infinity (the one at $-\infty$ and the other at $+\infty$) and thus has the topology $S^2\times R$ where $R$ is the real line. As time goes on, $\mathcal{S}_1$ evolves to $\mathcal{S}_2$. Since the line $\mathcal{S}_2$ connects the two origins $r=0$ (of two different coordinate
patches), it is compact and consequently it represents a closed three-dimentional surface with topology $S^3$, that of a three-sphere\footnote{Given a spacetime based on an $n$-dimensional manifold $M$ and an initial spacelike ($n - 1$)-dimensional hypersurface $\mathcal{S}_i$ and a final spacelike ($n - 1$)-dimensional hypersurface $\mathcal{S}_f$.  A topology change occurs if $\mathcal{S}_f$ is not diffeomorphic to $\mathcal{S}_i$~\cite{Dowker}.}.

It is generally believed that topology changes do not occur in classical physics and so they would be purely quantum phenomena~\cite{GH,Dowker}. No finalized theory of topology changes exists~\cite{Sorkin}, and it is reasonable to abandon the semi-classical approach in the Planck scale where quantum fluctuations become more important causing gravity to manifest itself in the form of an effective pressure that prevents matter from collapsing. It is admitted that the system (the collapsing black hole) makes a quantum jump with a change of spatial topology to avoid the creation of a singularity~\cite{Frolov}.

A subsequent investigation~\cite{topo2} has provided further clarifications on when the topology change occurs. It was argued that if the Dominant Energy Condition (DEC), $\rho_m\geq 0$ and $p_r$ and $p_t$ $\in [-c^2\rho_m,c^2\rho_m]$, is not violated (this implies the WEC is too not violated), the four necessary conditions of the Ref.~\cite{topo1}'s theorem are not sufficient to yield a topology change. Only if the DEC is violated but not the WEC, a topology change occurs. This was related to a sign change~\cite{topo2} of the curvature scalar~\eqref{cs1}, which is brought to the following form using $p_t=-c^2M(2\mathcal{D}+r\mathcal{D}')/2$ and $2\mathcal{D}=2\rho_m/M$:
\begin{equation}\label{cs2}
\mathcal{R}=-\frac{16\pi G}{c^4}~(c^2\rho_m-p_t)=\frac{16\pi G}{c^4}~(p_r+p_t).
\end{equation}
If the DEC is not violated, $\mathcal{R}<0$. Now, if the DEC is violated but not the WEC results in
\begin{equation}\label{cs3}
p_t>c^2\rho_m,
\end{equation}
and the sign of $\mathcal{R}$ changes from $-$ (in the region where the DEC is not violated) to $+$ (in the region where the DEC is violated).

\section{New distributions $\,\pmb{\mathcal{D}}$\label{secd}}
In this section we define new mathematical distributions $\mathcal{D}$ that mimic to a large extent the Dirac's one, then we discuss the special properties of the metric~\eqref{i4}.

\begin{table*}
\caption{{\footnotesize Some values of the distribution $\mathcal{D}_n(r,\theta)$~\eqref{d3} and of its cumulative distribution $\mathbb{D}_n(r,\theta)$~\eqref{d4} and~\eqref{d2}, expressed in terms of $\arctan$ and $\ln$ elementary functions. Here $x=r/\theta$~\eqref{d0b}. For the case $n=9/2$, $(-1)^{1/3}$ and $(-1)^{2/3}$ are complex numbers with $(-1)^{1/3}+(-1)^{2/3}=\sqrt{3}~{\rm i}$ and ${\rm i}^2=-1$ but the given expression of $\mathbb{D}_{9/2}(r,\theta)$ is real and is completely different from $-3 \sqrt{3} \ln (1+x^{3/2})/(2 \pi )$. The same remark applies to the case $n=18/5$. The corresponding metric solution~\eqref{i4} is given by~\eqref{d5}: $f=1-2MG\,\mathbb{D}_n/(c^2r)$. We provide the values of the coefficient $c_n$~\eqref{s5} and the extremal horizon $x_{\text{ext}}\equiv r_{\text{ext}}/\theta$~\eqref{s3b} in the fourth and fifth columns, respectively.}}\label{Tab1}
\begin{tabular}{|l|l|l|l|l|}
  \hline
  & & & & \vspace{-0.3cm} \\
  $n$ & $\mathcal{D}_n(r,\theta)$ & $\mathbb{D}_n(r,\theta)$ & $c_n$ & $x_{\text{ext}}$ \\
  & & & & \vspace{-0.3cm} \\
  \hline
  \hline
  & & & & \vspace{-0.3cm} \\
  $\frac{18}{5}$ & $\frac{9}{20 \pi ^2 \theta ^3 (x^{18/5}+1)}$ &
  $\frac{\rm i}{2 \pi } \ln \big(\frac{1-x^{3/5}{\rm i}}{1+x^{3/5}{\rm i}}\big)
  +\frac{1}{2 \pi } \big\{(-1)^{1/6} \big[\ln \big(\frac{1-(-1)^{1/6} x^{3/5}}{1+(-1)^{1/6}
  x^{3/5}}\big)+(-1)^{2/3} \ln \big(\frac{1-(-1)^{5/6} x^{3/5}}{1+(-1)^{5/6} x^{3/5}}\big)\big]\big\}$ & 0.842 & 1.882\\
  & & & & \vspace{-0.3cm} \\
  \hline
  & & & & \vspace{-0.3cm} \\
  $4$ & $\frac{1}{\sqrt{2} \pi ^2 \theta ^3 (x^4+1)}$ &
  $\frac{1}{2\pi }\big[2 \arctan (1+\sqrt{2} x)-2 \arctan (1-\sqrt{2} x)
  +\ln \big(\frac{x^2-\sqrt{2}x+1}{x^2+\sqrt{2}x+1}\big)\big]$ & 0.561 & 1.679 \\
  & & & & \vspace{-0.3cm} \\
  \hline
  & & & & \vspace{-0.3cm} \\
  $\frac{9}{2}$ & $\frac{9 \sqrt{3}}{16 \pi ^2 \theta ^3 (x^{9/2}+1)}$ &
  $-\frac{\sqrt{3}}{2\pi }\{\ln (1+x^{3/2})-(-1)^{1/3} \ln [1-(-1)^{1/3} x^{3/2}]
  +(-1)^{2/3} \ln [1+(-1)^{2/3} x^{3/2}]\}$ & 0.421 & 1.521 \\
  & & & & \vspace{-0.3cm} \\
  \hline
  & & & & \vspace{-0.3cm} \\
  $6$ & $\frac{3}{2 \pi ^2\theta ^3} \frac{1}{x^6+1}$ & $\frac{2}{\pi }\arctan (x^3)$ & 0.284 & 1.295 \\
  & & & & \vspace{-0.3cm} \\
  \hline
  & & & & \vspace{-0.3cm} \\
  $n$ & $\frac{n \sin \big(\frac{3 \pi }{n}\big)}{4 \pi ^2 \theta ^3 (x^n+1)}$ & $\frac{(-1)^{-3/n}}{\pi }\sin \Big(\frac{3 \pi }{n}\Big) \mathrm{B}\Big(-x^n;\,\frac{3}{n},0\Big)$ & $c_n$ & $x_{\text{ext}}$ \\
  \hline
\end{tabular}
\end{table*}

\subsection{Definition}
Let $\mathcal{A}(n,z)$ be the function defined by
\begin{equation}\label{d0}
\mathcal{A}(n,z)\equiv \int_0^z\frac{4\pi u^2}{u^n+1}~\dd u.
\end{equation}
If $z$ is real and $n>0$ (as we shall see later, we will require $n>3$ to ensure convergence of the integral), then
\begin{align}\label{d0c}
&\mathcal{A}(n,z)=\frac{(-1)^{-3/n}4\pi \mathrm{B}(-z^n;\,3/n,0)}{n},\\
&\mathcal{A}(n,\infty)=\frac{4\pi^2}{n\sin(3\pi/n)},\nn
\end{align}
where
\begin{equation}\label{d0d}
\mathrm{B}(z;\,a,b)\equiv \int_0^z t^{a-1}(1-t)^{b-1}\dd t,
\end{equation}
is the incomplete beta function\footnote{There are two notations for the incomplete beta function: $\mathrm{B}_{z}(a,b)$, used in~\cite{nist}, and $\mathrm{B}(z;\,a,b)$, used in~\cite{table} and http://mathworld.wolfram.com/IncompleteBetaFunction.html.}. One brings~\eqref{d0} to~\eqref{d0c} upon setting $t=-u^n$.

Using the new variable $x=r/\theta$~\eqref{d0b}, we define the distribution $\mathcal{D}_n(r,\theta)$ to be the function related to $\mathcal{A}(n,x)$ by
\begin{multline}\label{d1}
\mathcal{D}_n(r,\theta)\equiv \frac{1}{\mathcal{A}(n,\infty)\theta^3}~\frac{1}{x^n+1}\\
=\frac{\theta^{n-3}}{\mathcal{A}(n,\infty)}~\frac{1}{r^n+\theta^n},\; n>3,
\end{multline}
where $n$ is a real number and to ensure the convergence of the integral in~\eqref{i2} we have required $n>3$~\eqref{sr3c}. It is obvious from the definition that the distribution~\eqref{d1} reduces to the Dirac $\delta$ in the limit $\theta\to 0$. The cumulative distribution takes the form
\begin{equation}\label{d2}
\mathbb{D}_n(r,\theta)= \frac{\mathcal{A}(n,x)}{\mathcal{A}(n,\infty)},\quad n>3.
\end{equation}
Both $\mathcal{D}_n(r,\theta)$ and $\mathbb{D}_n(r,\theta)$ take the simplified expressions:
\begin{align}
\label{d3}&\mathcal{D}_n(r,\theta)=\frac{n \sin \big(\frac{3 \pi }{n}\big)}{4 \pi ^2 \theta ^3 (x^n+1)},\quad (n>3,\;x=r/\theta)\\
\label{d4}&\mathbb{D}_n(r,\theta)=\frac{(-1)^{-3/n}}{\pi }\sin \Big(\frac{3 \pi }{n}\Big) \mathrm{B}\Big(-x^n;\,\frac{3}{n},0\Big),
\end{align}
and the metric~\eqref{i4} reduces to
\begin{multline}\label{d5}
f(r)=1-\frac{2MG\,\mathbb{D}_n(r,\theta)}{c^2r}\\=1-\frac{2MG}{c^2r}~\frac{(-1)^{-3/n}}{\pi }\sin \Big(\frac{3 \pi }{n}\Big) \mathrm{B}\Big(-x^n;\,\frac{3}{n},0\Big).
\end{multline}
Using~\eqref{ip} it is easy to show that $\mathbb{D}_n(r,\theta)$ has one inflection point given by
\begin{equation}\label{ip2}
r=\Big(\frac{2}{n-2}\Big)^{1/n}\theta.
\end{equation}

There does not seem to be a special name given to the distributions of the form~\eqref{d1}. A distribution of the form \[\frac{1}{\pi (x^2+1)},\] is called the standard Cauchy\footnote{Its generalization~\cite{Cauchy2}, know as the generalized Cauchy distribution $f(z)$, is proportional to \[\frac{\sigma}{(\sigma^p+|z-\theta|^p)^{2/p}},\]
where $\theta$ is the location parameter, $\sigma$ is the scale parameter, and $p$ is the tail constant.}  distribution~\cite{Cauchy1}. It has been used to model dark haloes in spiral galaxies in the center and in the outer spatial regions~\cite{Cauchy3}; the model is widely accepted. An advantage in using the distribution~\eqref{d3} is that it depends on two parameters $(\theta,\,n)$. The distributions~\eqref{d3} and~\eqref{i1} have their denominators proportional to $\theta^3$, thus holding $\theta$ constant and varying $n$ one can generate a distribution~\eqref{d3} mimicking to a large extent the Dirac's one, as shown in Fig.~\ref{Fig2}, which is not possible with the Gaussian distribution~\eqref{i1}. Another advantage is that the cumulative distribution~\eqref{d4} can be brought to a closed-form (for all $n>3$) in terms of $\arctan$ and $\ln$ elementary functions. Table~\ref{Tab1} provides some distributions $\mathcal{D}_n(r,\theta)$ with their cumulative functions $\mathbb{D}_n(r,\theta)$.

\subsection{Special properties of the metric~\eqref{i4} and the physical scales}
In the previous section we discussed the general properties of the metric~\eqref{i4} that are independent of the special form of the distribution $\mathcal{D}(r,\theta)$. In this section we focus on other, rather specific, properties of~\eqref{i4} that result from the application of the distribution~\eqref{d3}: These are the properties of~\eqref{d5}.

\subsubsection{large $M$}
Introducing the parameters $x_h=r_h/\theta$ and $x_{\text{S}}=r_{\text{S}}/\theta$ we bring~\eqref{p1} to
\begin{equation}\label{ab1}
\frac{x_h}{x_{\text{S}}}=\mathbb{D}_n(x_h).
\end{equation}
For large $x$, the cumulative distribution is easily brought to the form
\begin{equation}\label{ab2}
\mathbb{D}_n\underset{(r \text{ large})}{\simeq} 1-\frac{\nu_n}{x^{n-3}}\quad\text{with}\quad
\nu_n\equiv \frac{n \sin \big(\frac{3 \pi }{n}\big)}{(n-3)\pi}.
\end{equation}
Using this in~\eqref{ab1} we solve it by iteration and obtain the outer horizon
\begin{equation}\label{ab3}
x_{h+}\underset{(M \text{ large})}{\simeq} x_{\text{S}}-\frac{\nu_n}{x_{\text{S}}^{n-4}}.
\end{equation}
The area of the outer horizon $A=\pi r_{h+}^2$ expands for large $r$, that is for large $M$, as
\begin{equation}\label{ab4}
A\underset{(M \text{ large})}{\simeq} \pi \theta^2 x_{\text{S}}^2\Big(1-\frac{2\nu_n}{x_{\text{S}}^{n-3}}\Big).
\end{equation}
The area spectrum follows the Bekenstein~\cite{particles} law
\begin{equation}\label{ab5}
A=bN \quad\text{with}\quad b\equiv 4\ell_{\text{P}}^2\ln 2\quad\text{and}\quad N\in\mathbb{N}^+.
\end{equation}
Here $N$ is a positive integer and $\ell_{\text{P}}=\sqrt{\hbar G/c^3}\simeq 1.616\times 10^{-35}\text{ m}$ is the Planck length. If the black hole emits a quanta, that is, if $N$ changes by 1 ($\dd N=-1$), this yields a change in the mass parameter $M$ given by the first order approximation
\begin{equation}\label{ab6}
    |\dd M|\simeq \frac{bM}{2\pi\theta^2x_{\text{S}}^2}=\frac{bc^4}{8\pi MG^2}\ll 1,
\end{equation}
which is independent of $\theta$. Since $b\ll 1$ and $M$ is supposed large, this implies that the change in $M$ or the mass loss is almost continuous.

\subsubsection{Extremal horizon}
The value of $x_{\text{ext}}$ is solution to~\eqref{p3}, which takes the form
\begin{equation}\label{s3b}
\frac{x_{\text{ext}}^3}{x_{\text{ext}}^n+1}=\int_{0}^{x_{\text{ext}}}\frac{u^2}{u^n+1}~\dd u,
\end{equation}
The solution of which yields an $x_{\text{ext}}$ independent of the mass $M$.

\subsubsection{Temperature}
With $\mathcal{D}$ given by~\eqref{d3} the expression of $T$~\eqref{t2} reduces to
\begin{equation}\label{mt0}
T(r_{h+})=\frac{\hbar c}{4\pi k_{\text{B}}r_{h+}}\bigg(1-\frac{n}{\pi}~\sin \Big(\frac{3 \pi }{n}\Big) \frac{x_{\text{S}}x_{h+}}{x_{h+}^n+1}\bigg),
\end{equation}
where $x_{\text{S}}=x_h/\mathbb{D}_n(x_h)$~\eqref{ab1}. For $\mathcal{D}$ given by~\eqref{d3} it is not easy to determine the value of the outer horizon $\tilde{r}_{h+}$ that yields a maximum temperature, while for a Gaussian distribution~\eqref{i1} $\tilde{r}_{h+}=\sqrt{2}~\theta$.

\subsubsection{Topology change}
The condition $p_t>c^2\rho_m$~\eqref{cs3} is brought to $r|\partial_r\rho_m|>4\rho_m$ or, equivalently, to $r|\mathcal{D}'|>4\mathcal{D}$. Using the expression~\eqref{d3} of $\mathcal{D}$ we arrive at
\begin{equation}\label{cs4}
(n-4)r^n>4\theta^n.
\end{equation}
Thus, for $3<n\leq 4$ there is no topology change and $\mathcal{R}<0$ for all $r$. For $n>4$, the DEC is violated but not the WEC and a toplogy change occurs along with a sign change of $\mathcal{R}$ where it becomes positive for
\begin{equation}\label{cs5}
r>r_{\star}\equiv \Big(\frac{4}{n-4}\Big)^{1/n}\theta .
\end{equation}

\subsubsection{From Planckian scale to stellar scale\label{secps}}
For large $M$, the metric~\eqref{d5} expands as
\begin{equation}\label{ab7}
f\underset{(M \text{ large})}{\simeq}1-\frac{x_{\text{S}}}{x}+\frac{\nu_n x_{\text{S}}}{x^{n-2}}.
\end{equation}
The inner horizon $r_{h-}$ is obtained upon solving the algebraic equation:
\begin{equation}\label{ab8}
x_{h-}^{n-2}-x_{\text{S}}x_{h-}^{n-3}+\nu_nx_{\text{S}}\simeq 0.
\end{equation}
For instance, for $n=5$ we obtain
\begin{equation}\label{ab9}
r_{h-}\underset{(M \text{ large})}{\simeq}\sqrt{\nu_5}~\theta+\frac{\nu_5\theta^2}{2r_{\text{S}}},\;\quad \Big(x_{h-}\simeq\sqrt{\nu_5}+\frac{\nu_5}{2x_{\text{S}}}\Big),
\end{equation}
where $\nu_5=5\sin(3\pi/5)/(2\pi)\approx 582/769\simeq 0.756827$.

The metric~\eqref{ab7} is a quintessence-like metric. Knowing that a Gaussian distribution~\eqref{i1} does not accurately describe~\cite{inspired} the galaxies rotation curves~\cite{rotation} as does, for instance, the pseudoisothermal model~\cite{Cauchy3}. This shows another advantage in using the distributions~\eqref{d3}, for they can model dark matter distributions better than a Gaussian distribution and provide best fits compared to the standard models~\cite{Cauchy3,Cauchy4,Cauchy5}. From this point of view $\theta$ is of the order of the stellar or core radius.

The metric~\eqref{ab7} is also of the form of a quantum-corrected Schwarzschild black hole~\cite{qcbh1,qcbh2,qcbh3}. It is of the same form as Eq.~(3) of Ref.~\cite{qcbh2} provided we take
\begin{equation}\label{ab10}
n=5\quad\text{and}\quad\nu_5\theta^2\equiv \gamma\ell_{\text{P}}^2.
\end{equation}
With these identifications the outer~\eqref{ab3} and inner~\eqref{ab9} horizons coincide with the solutions given in Eqs.~(41) and~(42) of Ref.~\cite{qcbh2}. The parameters $\nu_5$ and $\gamma>0$~\cite{qcbh2} being of the order of unity, we see that from this point of view $\theta$ is of the order of the Planck length.

Considering regular particle-sized black holes with a Gaussian mass distribution~\eqref{i1}, it was argued in Ref.~\cite{SS3} that a qualitative realization of the UV self-completeness of quantum gravity could be achieved taking $\theta$ of the order of the Compton wavelength of a particle of mass $M$: $\theta\sim 1/M$. This scheme can be easily realized using our model for mass distributions given by~\eqref{d3}.

Thus, the parameter $\theta$ provides three length scales of application:
\begin{enumerate}
  \item A cosmological scale where $\theta$ is of the order of the stellar or core radius;
  \item A subatomic scale where $\theta$ is of the order of the Compton wavelength of a particle of mass $M$;
  \item A Planckian scale for describing quantum-corrected black holes.
\end{enumerate}
For black hole or particle-like solutions there are, however, other means by which one may constrain the values of $\theta$, as we shall discuss in the remaining sections.

\subsubsection{Radius of fuzzy matter distributions}
It is straightforward to show that the transverse pressure~\eqref{set1} is up to a constant factor given by
\begin{equation}
p_t\propto \frac{(n-2)r^n-2\theta^n}{2(r^n+\theta^n)^2},
\end{equation}
for all $r$. This vanishes in the limit $r\to\infty$. It is negative for $0\leq r<r_0$, null for $r=r_0$, and positive for $r>r_0$ where
\begin{equation}
r_0\equiv \Big(\frac{2}{n-2}\Big)^{1/n}\theta,
\end{equation}
which is just the point where $\mathbb{D}_n(r,\theta)$ has its inflection point~\eqref{ip2}. First note that $r_{\star}>r_0$~\eqref{cs5}. One may call the value $r_0$ the distributional radius of the black hole or that of the galaxy. It is a measure of the distance beyond which the effects of vacuum due to the fuzzy distribution of matter tend to be neglected. For a Gaussian distribution~\eqref{i1}, $r_0=\infty$, that is, the tangential pressure is negative for the whole range of the radial coordinate. One sees that the distributions~\eqref{d3} provide more realistic models for describing fuzzy matter distributions or galactic dark matter halos.

For black hole solutions one may constrain the values of $\theta$ upon requiring that all the fuzzy matter distribution be confined within the inner horizon. This allows one to describe classically the geometry outside the event horizon.

\section{Limiting values\label{seclc}}
Using~\eqref{d3} we bring~\eqref{ic} to
\begin{equation}\label{l1}
    \Lambda=\frac{MG}{\pi c^2\theta ^3}~n \sin \Big(\frac{3 \pi }{n}\Big).
\end{equation}
If we assume that the smallest value of $\Lambda$ is the cosmological constant $\Lambda_{\text{csm}}$, this yields the maximum value for $\theta$
\begin{equation}\label{l2}
\theta_{\text{max}}{}^3=3\frac{\ell_{\text{P}}}{\Lambda_{\text{csm}}}~\frac{M}{m_{\text{P}}},
\end{equation}
where we replaced $n\sin(3 \pi /n)$ by its upper bound $3\pi$. Here $m_{\text{P}}=\sqrt{\hbar c/G}\simeq 2.177\times 10^{-8}\text{ kg}$ is the Planck mass.

An upper bound for $\Lambda$ could be set requiring the width $\theta$ to be of the order of the Compton wavelength $\hbar/(Mc)$ for the mass $M$, which expresses the inability to localize a single particle in a region of size $\hbar/(Mc)$. We obtain
\begin{equation}\label{l3}
\Lambda_{\text{max1}}=n \sin \Big(\frac{3 \pi }{n}\Big)\frac{M^4}{\pi m_{\text{P}}^4}~\ell_{\text{P}}^{-2}< 3\frac{M^4}{m_{\text{P}}^4}~\ell_{\text{P}}^{-2},
\end{equation}
where we replaced $n\sin(3 \pi /n)$ by its upper bound $3\pi$.

For describing quantum-corrected black holes $\theta$ could be of the order of the Planck length. For these holes, an upper bound for $\Lambda$ is rather
\begin{equation}\label{l4}
\Lambda_{\text{max2}}\approx 3\frac{M}{m_{\text{P}}}~\ell_{\text{P}}^{-2}.
\end{equation}

\begin{table}
	\caption{{\footnotesize The masses are in solar mass units and radii are in solar radius units. The data has been reported in Refs.~\cite{data5}, \cite{data6}, and \cite{data7}.}}\label{Tab2}
\begin{tabular}{lll}
  \hline
  & & \vspace{-0.3cm} \\
  Star\hspace{5mm} &\hspace{5mm} $M$ ($\times$ M$_{\odot}$)\hspace{5mm} &\hspace{5mm} $a$ ($\times$ R$_{\odot}$) \\
  & & \vspace{-0.3cm} \\
  \hline\hline
  & & \vspace{-0.3cm} \\
  Sirius B\hspace{5mm} &\hspace{5mm} 1.034\hspace{5mm} &\hspace{5mm} 0.0084 \\
  & & \vspace{-0.3cm} \\
  & & \vspace{-0.3cm} \\
  Sun\hspace{5mm} &\hspace{5mm} 1\hspace{5mm} &\hspace{5mm} 1 \\
  & & \vspace{-0.3cm} \\
  & & \vspace{-0.3cm} \\
  Procyon B\hspace{5mm} &\hspace{5mm} 0.604\hspace{5mm} &\hspace{5mm} 0.0096 \\
  & & \vspace{-0.3cm} \\
  & & \vspace{-0.3cm} \\
  40 Eri B\hspace{5mm} &\hspace{5mm} 0.501\hspace{5mm} &\hspace{5mm} 0.0136 \\
  & & \vspace{-0.3cm} \\
  & & \vspace{-0.3cm} \\
  EG 50\hspace{5mm} &\hspace{5mm} 0.50\hspace{5mm} &\hspace{5mm} 0.0104 \\
  & & \vspace{-0.3cm} \\
  & & \vspace{-0.3cm} \\
  GD 140\hspace{5mm} &\hspace{5mm} 0.79\hspace{5mm} &\hspace{5mm} 0.0085 \\
  & & \vspace{-0.3cm} \\
  & & \vspace{-0.3cm} \\
  CD-38 10980\hspace{5mm} &\hspace{5mm} 0.74\hspace{5mm} &\hspace{5mm} 0.01245 \\
  & & \vspace{-0.3cm} \\
  & & \vspace{-0.3cm} \\
  W485A\hspace{5mm} &\hspace{5mm} 0.59\hspace{5mm} &\hspace{5mm} 0.0150 \\
  & & \vspace{-0.3cm} \\
  & & \vspace{-0.3cm} \\
  G154-B5B\hspace{5mm} &\hspace{5mm} 0.46\hspace{5mm} &\hspace{5mm} 0.0129 \\
  & & \vspace{-0.3cm} \\
  & & \vspace{-0.3cm} \\
  LP 347-6\hspace{5mm} &\hspace{5mm} 0.56\hspace{5mm} &\hspace{5mm} 0.0124 \\
  & & \vspace{-0.3cm} \\
  & & \vspace{-0.3cm} \\
  G181-B5B\hspace{5mm} &\hspace{5mm} 0.54\hspace{5mm} &\hspace{5mm} 0.0125 \\
  & & \vspace{-0.3cm} \\
  & & \vspace{-0.3cm} \\
  WD1550+130\hspace{5mm} &\hspace{5mm} 0.535\hspace{5mm} &\hspace{5mm} 0.0211 \\
  & & \vspace{-0.3cm} \\
  & & \vspace{-0.3cm} \\
  Stein 2051B\hspace{5mm} &\hspace{5mm} 0.48\hspace{5mm} &\hspace{5mm} 0.0111 \\
  & & \vspace{-0.3cm} \\
  & & \vspace{-0.3cm} \\
  G107-70AB\hspace{5mm} &\hspace{5mm} 0.65\hspace{5mm} &\hspace{5mm} 0.0127 \\
  & & \vspace{-0.3cm} \\
  & & \vspace{-0.3cm} \\
  L268-92\hspace{5mm} &\hspace{5mm} 0.70\hspace{5mm} &\hspace{5mm} 0.0149 \\
  & & \vspace{-0.3cm} \\
  & & \vspace{-0.3cm} \\
  G156-64\hspace{5mm} &\hspace{5mm} 0.59\hspace{5mm} &\hspace{5mm} 0.0110 \\
  \hline
\end{tabular}
\end{table}

\section{New metric solutions\label{secs}}
Selecting the simplest solution given in Table~\ref{Tab1} we are led to the following regular metric
\begin{equation}\label{s1}
f(r)=1-\frac{4 M G}{\pi  c^2 r}~\arctan \Big(\frac{r^3}{\theta ^3}\Big).
\end{equation}
This is a substitute to the singular Schwarzschild metric resulting from the substitution rule~\eqref{sr1}
\begin{equation}\label{s2}
\frac{\delta (r)}{2 \pi  r^2}\to \frac{3 \theta ^3}{2 \pi ^2} \frac{1}{r^6+\theta ^6}=\frac{3}{2 \pi ^2 \theta ^3 (x^6+1)},
\end{equation}
where we have replaced the Gaussian distribution by~\eqref{d3} taking $n=6$. The corresponding continuous mass density $\rho _m(r)$ and mass $m(r)$ within a sphere of radius $r$ are given by
\begin{equation}\label{s3}
\rho _m(r)=\frac{3 \theta ^3}{2 \pi ^2} \frac{M}{r^6+\theta ^6},\quad m(r)=\frac{2 M}{\pi }~\arctan \Big(\frac{r^3}{\theta ^3}\Big).
\end{equation}

Plots of~\eqref{s1} are shown in Fig.~\ref{Fig3} for different values of $MG/(\pi c^2\theta)$. For large values of $M$ the solution is a double-horizon black hole and for smaller values of $M$ the solution is a quantum particle or a regular non-black-hole solution. For some intermediate value of $M=M_{\text{ext}}$ such that $M_{\text{ext}}G/(\pi c^2\theta)\simeq 0.284$ ($x_{\text{S}}\simeq 2\pi\times 0.284$) the two horizons merge forming one extremal horizon at $r_{\text{ext}}\simeq 1.295~\theta$. The value of $x_{\text{ext}}$ is solution to~\eqref{s3b}.

This solution models a regular noncommutative black hole where the effects of noncommutativity of coordinates are phenomenologically played by a smeared, extended, mass distribution. For larger values of $n$, the mass distribution~\eqref{d3}, being almost a step function (see Fig.~\ref{Fig2}), is more confined in a region around the black hole and the solution represents a classical black hole. For smaller values of $n$ the distribution~\eqref{d3} is more extended, like a Gaussian distribution, and the solution represents a semi-classical black hole.

One may ask: What is the upper limit of the ratio $M/\theta$, where $\theta$ is a measure of the extent of matter, that prevents the occurrence of horizons? The answer is as follows.

For modeling dark matter halos one may apply the mass distribution~\eqref{s3} to halos with stellar radius $a$ and mass $M$ such that
\begin{equation}\label{s4}
M<\frac{c_6\pi c^2a}{G}\quad\text{ and }\quad c_6=0.284,
\end{equation}
so as to avoid the formation of black-hole dark matter halos. The mass within a sphere of radius $r$ is given by~\eqref{s3} on replacing $\theta$ by $a$.

Such an upper limit on $M$ is not absolute, that is, larger dark matter halos are modeled by the distribution~\eqref{d3} taking $n<6$. This will set another upper limit for the mass for such halos similar to~\eqref{s4} with a new coefficient $c_n$ larger than, but remains of the same order of $0.284$ for the values of $n$ considered in Table~\ref{Tab1}:
\begin{equation}\label{s5}
M<\frac{c_n\pi c^2a}{G}.
\end{equation}
Table~\ref{Tab1} provides some values of $c_n$. This allows us to claim that the mass-to-radius ratio of dark matter halos is of the order of
\begin{equation}\label{s6}
\frac{M}{a}\lesssim\mu\equiv\frac{\pi c^2}{G}=\frac{\pi c}{\hbar}~m_{\text{P}}{}^2=4.23126\times 10^{27}\text{ kg/m}.
\end{equation}
This upper limit is at least satisfied by the dwarf galaxies with stellar radii 10-30 kpc as can be seen from the dark matter profiles~\cite{Cauchy5} derived from the data of rotation curves of the DDO 154, DDO 105, NGC 3109, and DDO 170 spiral galaxies reported in Refs.~\cite{data1}, \cite{data2}, \cite{data3}, and~\cite{data4}, respectively. The scaling empirical Eq. (3) of Ref.~\cite{Cauchy5} correlates the dark matter mass $M$ inside a sphere of radius $a$ where the ratio $M/a$ remains of the order of $10^{20}\text{ kg/m}$.

For modeling stars one may apply the mass distribution~\eqref{s3} to stars with radius $a$ and mass $M$ such that~\eqref{s4} is satisfied so as to avoid the formation of a black hole. This is justified since the graph of $\mathcal{D}_6$, shown in Fig.~\ref{Fig2}, is almost similar to that of a step function; the mass distribution vanishes almost identically for $r>\theta$, and vanishes faster than a Gaussian distribution in the vicinity of $r\gtrapprox\theta$. For lighter stars we may take $n>6$ in~\eqref{d3} so that the mass remains confined inside the sphere of radius $\theta$ (the radius of the star). We reach the same conclusion as before, in that, the ratio $M/a$ remains bounded from above by the constant $\mu$ defined in~\eqref{s6}. For the stars of Table~\ref{Tab2}, the data of which has been reported in Refs.~\cite{data5}, \cite{data6}, and \cite{data7}, the ratio $M/a\sim 10^{23}\text{ kg/m}<\mu$.

For modeling elementary particles we take $\theta$ to be of the order of the (reduced) Compton wavelength, $a=\theta=\hbar/(Mc)$, and $n\geq 6$ so that the shape of the mass distribution be that of a step function. In this case $\mathcal{D}_n$ depends on the mass of the particle via $\theta$. The condition~\eqref{s5} ensuring the absence of horizons reduces to
\begin{equation}\label{s7}
M^2<\pi m_{\text{P}}{}^2,
\end{equation}
where we have dropped $c_n$. This is the well known property stating that the masses of elementary particles are much smaller than the Planck mass $m_{\text{P}}$.

We draw the general conclusion that \emph{any mass distribution of extent $\theta$ and mass $M$ is exempt of, or freed from, horizons if}
\begin{equation}\label{s8}
\frac{M}{\theta}\lesssim\mu=\frac{\pi c}{\hbar}~m_{\text{P}}{}^2.
\end{equation}
The largeness of the constant $\mu$~\eqref{s8} is behind the difficulty in manufacturing laboratory black holes by compressing solids. To achieve that one should reduce the size of the solid, with given mass $M$, to below $M/\mu$.

This may apply to the whole universe itself: If the ratio (mass of the universe/extent of the universe) is larger than $\mu$, we may be living inside a two or multi-horizon black hole, most likely inside the inner horizon. Otherwise, the space around us is freed from horizons.

Can the metric~\eqref{s1}, which corresponds to $n=6$, describe a quantum-corrected Schwarzschild black hole? In Sec.~\ref{secps} we have seen that such a black hole can be described by a mass distribution~\eqref{d3} provided we take $n=5$. The metric expansion~\eqref{ab7} with $n=5$,
\begin{equation}\label{s9}
    f\simeq 1-\frac{r_{\text{S}}}{r}+\frac{\gamma\ell_{\text{P}}^2 r_{\text{S}}}{r^{3}},
\end{equation}
which describes a quantum-corrected Schwarzschild black hole, has been first derived in~\cite{qcbh1} upon evaluating the self-energy insertion tensor (SEIT)~\cite{qcbh4} due to the inclusion of a single-closed loop, which is a quantum correction. The finite piece of the SEIT contains some arbitrary parameters while the infinite piece is supposed to be canceled by appropriate counter-terms in the Lagrangian. However, it is all possible that these canceling counter-terms may alter the values of the parameters in the finite piece of the SEIT causing the final expression of the metric~\eqref{s9} to include, say, a term proportional to $1/r^4$ or other powers of $1/r$ instead of a term proportional to $1/r^3$.

\section{Cumulative distributions with many inflection points: Core--multi-shell regular black holes  \label{cdip}}
All we have dealt with in the previous sections concerned cumulative distributions with one inflection point. On large scales, mass distributions may not be central; rather, spread onto concentric \emph{extended} shells or accretion disks with voids in between. The location of the voids are nearly coincident with the inflection points of the cumulative distribution. As we have seen earlier, such mass distributions with $2k+1$ inflection points may have up to $2(k+1)$ horizons~\eqref{p2}.

We present two ways to construct such multi-horizon solutions. In these constructions we take the mass distribution~\eqref{s3} as a prototype. Notice that
\begin{equation}\label{w1}
    \int_0^r\frac{u^{i-1}}{u^{2i}+\theta^{2i}}~\dd u=\frac{1}{i\theta^i}~\arctan\Big(\frac{r^i}{\theta ^i}\Big),\qquad (i \text{ integer}),
\end{equation}
so, in order to obtain simple solutions, we choose the mass distribution of the form
\begin{equation}\label{w2}
\rho_m(r)=M\mathcal{D}(r,\theta)\equiv\sum_{i=1}\frac{(2+i)\theta_i^{2+i}M_i}{2\pi^2}~\frac{r^{i-1}}{r^{2(2+i)}+\theta_i^{2(2+i)}},
\end{equation}
where $M=\sum_{i=1}M_i$ and $\theta=(\theta_1,\,\theta_2,\,\cdots,\theta_i,\,\cdots)$. The coefficients have been chosen so that the integral of each term on the whole range of $r$ is $M_i$, $M_1$ being the mass of the central core and $M_i$ with $i\geq 2$ are the masses of shells. This yields
\begin{equation}\label{w3}
\mathbb{D}(r,\theta)=\frac{2}{\pi}~\sum_{i=1}M_i\arctan\Big(\frac{r^{2+i}}{\theta_i^{2+i}}\Big),
\end{equation}
which represents a core--multi-shell regular black hole.

\subsection{Core--shell regular black hole}
Keeping the first two terms in~\eqref{w2} and~\eqref{w3} we obtain a core-shell regular black hole,
\begin{equation}\label{w3b}
\rho_m(r)=\frac{3\theta_1^3M_1}{2\pi^2}~\frac{1}{r^6+\theta_1^6}
+\frac{4\theta_2^4M_2}{2\pi^2}~\frac{r}{r^8+\theta_2^8}
\end{equation}
\begin{equation}\label{w4}
\mathbb{D}(r,\theta)=\frac{2}{\pi}~\sum_{i=1}^2 M_i\arctan\Big(\frac{r^{2+i}}{\theta_i^{2+i}}\Big),
\end{equation}
where the graph of $\mathbb{D}(r,\theta)$ is shown in the second plot from the left of Fig.~\ref{Fig1} taking $M_2=2/3=2M_1$ ($M=1$) and $\theta_2=5=8\theta_1$.

\subsection{Core--two-shell and two-shell regular black holes}
Keeping the first three terms in~\eqref{w2} and~\eqref{w3} we obtain a core--two-shell regular black hole,
\begin{equation}\label{w5}
\rho_m(r)=\frac{3\theta_1^3M_1}{2\pi^2}~\frac{1}{r^6+\theta_1^6}
+\frac{4\theta_2^4M_2}{2\pi^2}~\frac{r}{r^8+\theta_2^8}+\frac{5\theta_3^5M_3}{2\pi^2}~\frac{r^2}{r^{10}+\theta_3^{10}}
\end{equation}
\begin{equation}\label{w6}
\mathbb{D}(r,\theta)=\frac{2}{\pi}~\sum_{i=1}^3 M_i\arctan\Big(\frac{r^{2+i}}{\theta_i^{2+i}}\Big),
\end{equation}
where the graph of $\mathbb{D}(r,\theta)$ is shown in the third plot from the left of Fig.~\ref{Fig1} taking $M_1=0.3$, $M_2=0.8$, $M_3=1.2$ ($M=2.3$), $\theta_1=1$, $\theta_2=4$, and $\theta_3=9.5$.

Now, keeping for instance the second and third terms in~\eqref{w2} and~\eqref{w3} we obtain a two-shell regular black hole with total mass $M=M_2+M_3$ and whose mass density and cumulative distribution are given by
\begin{equation}
\rho_m(r)=\frac{4\theta_2^4M_2}{2\pi^2}~\frac{r}{r^8+\theta_2^8}+\frac{5\theta_3^5M_3}{2\pi^2}~\frac{r^2}{r^{10}+\theta_3^{10}}
\end{equation}
\begin{equation}
\mathbb{D}(r,\theta)=\frac{2}{\pi}~\sum_{i=2}^3 M_i\arctan\Big(\frac{r^{2+i}}{\theta_i^{2+i}}\Big).
\end{equation}

In the same manner we can obtain a multi-shell regular black hole.

\subsection{Another core--shell regular black hole}
One may obtain core--multi-shell regular black holes upon shifting, horizontally, the graph of the mass distribution~\eqref{s3}. This yields a piecewise solution where the mass distribution of~\eqref{s3} represents the core and the shifted graph represents the shell. We choose $\rho_m(r)=M\mathcal{D}(r,\theta)$ such that
\begin{equation}\label{w7}
\mathcal{D}(r,\theta)=\left\{\begin{array}{ll}
\mathcal{D}_1(r,\theta_1)=\frac{c_1}{r^6+\theta_1^6}, & \hbox{$0\leq r\leq a$;} \\
\mathcal{D}_2(r,\theta_2)=\frac{c_2r^2+br+c}{(r-a)^6+\theta_2^6}, & \hbox{$r>a$}
                \end{array}
            \right..
\end{equation}
We can determine $b$ and $c$ in terms of the other parameters on imposing the continuity of $\mathcal{D}(r,\theta)$ and of its $r$ derivative, $\mathcal{D}'(r,\theta)$, at $r=a$ (no jump discontinuities at $r=a$ so that the scalar invariants $R$ and $R_{\al\bt\mu\nu}R^{\al\bt\mu\nu}$ remain finite). In order to fix the other two constants we require that
\begin{align}
&\int_0^a 4\pi u^2\mathcal{D}_1(u,\theta_1)\dd u=\frac{M_1}{M},\nn\\
&\int_a^{\infty} 4\pi u^2\mathcal{D}_2(u,\theta_2)\dd u=\frac{M_2}{M},\\
&M=M_1+M_2\nn ,
\end{align}
where $M_1$ is the mass of the central core and $M_2$ is that of the shell. For instance, if we take
\begin{equation}\label{w8}
\theta_1=\theta_2=a\equiv \theta,
\end{equation}
we find
\begin{align}
&c_1=\frac{3 M_1 \theta ^3}{\pi ^2 M},\;c_2=\frac{3 [3 M+(6+8 \sqrt{3}) M_1] \theta }{2 (9+4 \sqrt{3}) \pi ^2 M},\nn\\
\label{w9}&c=\frac{9[M+2 (7+4 \sqrt{3}) M_1] \theta ^3}{2 (9+4 \sqrt{3}) \pi ^2 M},\\
&b=-\frac{3 [6 M+(39+28 \sqrt{3})M_1] \theta ^2}{2 (9+4 \sqrt{3}) \pi ^2 M}\nn .
\end{align}
A plot of the cumulative distribution of~\eqref{w7}, with its parameters as given in~\eqref{w8} and~\eqref{w9}, is shown in the right-most plot of Fig.~\ref{Fig1} taking $\theta=1$, $M_1=M/7$, and $M_2=6M/7$.

The thermodynamics of these core--multi-shell regular black holes deserves a special treatment that is out of the scope of this paper.

\section{Conclusion \label{secc}}
A way to describe phenomenologically dark matter halos, stars, effects of noncommutativity and quantum corrections to stellar objects is to model them by extended mass distributions.

A variety of such mass distributions as well as charge distributions~\cite{comment,Weibull} have been put forward for the sole purpose mentioned above. To the best of our knowledge only a Gaussian mass distribution has received a two-fold application: Constructing noncommutative black holes and describing dark matter halos.

We have discussed the generic properties of these mass distributions. Their resulting metric solutions all have a de Sitter behavior near the origin, finite scalar invariants, and finite temperature if they describe black holes. In the latter case, the evaporation processes are marked by the finiteness of the temperature which first increases to a maximum value then decreases to absolute zero at the end of the process, contrary to the Schwarzschild case where the temperature unceasingly increases to infinity during the process of evaporation.

Then we have specialized to a new class of mass distributions. We have defined and used step-like mass distributions. Being dependent on two independent parameters, these distributions are multi-fold and they apply to a variety of physical configurations ranging from noncommutative black holes, quantum-corrected black holes to stars and dark matter halos depending on different scaling values of one of the two parameters.

The resulting regular metric solution is always given in closed form in terms of the arctan and ln elementary functions. For linear mass densities exceeding $\pi c^2/G$, the geometry is that of a two-horizon regular black hole; otherwise the geometry is freed from horizons and describes a regular non-black-hole configuration that could be a quantum particle, a star, a dark matter halo, the whole universe, or a compressed quantum solid.

Core--multi-shell and multi-shell regular black holes were also the subject of this work. We have presented two different ways to construct these objects which represent final stages of matter collapse into regular configurations.


\section*{Appendix: Equation yielding the extremal horizon and maximum temperature for generic mass distribution\label{secaa}}
\renewcommand{\theequation}{A.\arabic{equation}}
\setcounter{equation}{0}
Using the new variable $x=r/\theta$ we bring~\eqref{i3b} and~\eqref{p1} to
\begin{align}
\label{A1}&\mathbb{D}(x)= 4\pi \theta^3\int_0^{x}\mathcal{D}(u)u^2\dd u,\\
&\frac{x}{x_{\text{S}}}=\mathbb{D}(x).
\end{align}

Excluding the point $x=0$, the line $y=x/x_{\text{S}}$ is tangent to the curve $y=\mathbb{D}(x)$ at the only intersection point $x_{\text{ext}}$ if $\mathbb{D}$ has one inflection point; if $\mathbb{D}$ has $2k+1$ inflection points~\eqref{p2}, there could be up to $k+1$ tangential intersection points $x_{\text{ext}}$ satisfying the system
\begin{align}
\label{A3}&\frac{x_{\text{ext}}}{x_{\text{S}}}=\mathbb{D}(x_{\text{ext}}),\\
\label{A4}&\frac{1}{x_{\text{S}}}=\partial_x\mathbb{D}(x)\big|_{x=x_{\text{ext}}}.
\end{align}
From~\eqref{A1} we obtain $\partial_x\mathbb{D}(x)\big|_{x=x_{\text{ext}}}=4\pi \theta^3\mathcal{D}(x_{\text{ext}})x_{\text{ext}}^2$ and the system~\eqref{A3}-\eqref{A4} reduces to
\begin{align}
\label{A5}&\frac{x_{\text{ext}}}{x_{\text{S}}}=4\pi \theta^3\int_0^{x_{\text{ext}}}\mathcal{D}(u)u^2\dd u,\\
\label{A6}&\frac{1}{x_{\text{S}}}=4\pi \theta^3\mathcal{D}(x_{\text{ext}})x_{\text{ext}}^2,
\end{align}
which upon eliminating $x_{\text{S}}$ yields the integral-algebraic Eq.~\eqref{p3}
\begin{equation*}
\mathcal{D}(x_{\text{ext}})x_{\text{ext}}^3=\int_{0}^{x_{\text{ext}}}\mathcal{D}(u)u^2\dd u.
\end{equation*}

With $\partial_x\mathbb{D}(x)\big|_{x=x_{\text{ext}}}=4\pi \theta^3\mathcal{D}(x_{\text{ext}})x_{\text{ext}}^2$, Eq.~\eqref{A6} reduces to~\eqref{p3b}.

Note that~\eqref{A6} may be arranged as
\begin{equation*}
1-4\pi x_{\text{S}}\theta^3\mathcal{D}(x_{\text{ext}})x_{\text{ext}}^2=1-4\pi r_{\text{S}}\mathcal{D}(r_{\text{ext}})r_{\text{ext}}{}^2=0,
\end{equation*}
which implies that the temperature~\eqref{t2} vanishes at $r_{\text{ext}}$.

The temperature $T$ is proportional to
\begin{equation}\label{A7}
\partial_r\ln\Big(\frac{r}{\mathbb{D}}\Big)\bigg|_{r=r_{h+}},
\end{equation}
yielding
\begin{equation}\label{A8}
\partial_{r_{h+}}T\propto -\frac{1}{r_{h+}^2}-\partial_r\Big(\frac{4\pi r^2\mathcal{D}}{\mathbb{D}}\Big)\bigg|_{r=r_{h+}},
\end{equation}
where we have used $\mathbb{D}'=4\pi r^2\mathcal{D}$. The equation $\partial_{r_{h+}}T\big|_{r_{h+}=\tilde{r}_{h+}}=0$ yields~\eqref{t4}.



\end{document}